\documentclass[useAMS,usenatbib]{mn2e}
\usepackage{epsfig}
\usepackage{color} 
\usepackage{natbib}
\usepackage{deluxetable}
\usepackage{amssymb}
\usepackage{rotating}
\usepackage{graphicx}
\usepackage[colorlinks=true,citecolor=blue]{hyperref}
\usepackage{mathpazo}
\usepackage{graphicx}
\usepackage{footnote}
\usepackage{longtable}

\title[INOV of $\gamma$-ray detected NLSy1s]
{Intra-night optical variability of $\gamma-$ray detected narrow-line Seyfert1 galaxies}
\author[Ojha et al. ]{Vineet Ojha$^{1}$\thanks{E-mail: vineet@aries.res.in, vineetojhabhu@gmail.com
  }, Hum Chand$^{2,~1}$, Gopal-Krishna$^{3,~1}$   \\
$^{1}$Aryabhatta Research Institute of Observational Sciences (ARIES), Manora Peak, Nainital 
  $-$ 263002, India.\\
$^{2}$Department of Physics and Astronomical Sciences, Central University of Himachal Pradesh (CUHP), Dharamshala$-$176215, India\\  
$^{3}$UM-DAE Centre for Excellence in Basic Sciences, Vidyanagari, Mumbai-400098, India.\\}

\begin{document}
\date{Accepted ---. Received ---; in original form ---}

\pagerange{\pageref{firstpage}--\pageref{lastpage}} \pubyear{2017}

\maketitle

\label{firstpage}
\begin{abstract}\\
We report the first attempt to systematically characterise intra-night optical variability (INOV) of the rare and enigmatic subset of  Narrow-Line Seyfert1 galaxies (NLSy1s), which is marked by detection in the $\gamma$-ray band and is therefore endowed with Doppler boosted relativistic jets, like blazars. However, the central engines in these two types of AGN are thought to operate in different regimes of accretion rate. Our INOV search in a fairly large and unbiased sample of 15 $\gamma$-ray NLSy1s was conducted in 36 monitoring sessions, each lasting $\geq$ 3 hrs. In our analysis, special care has been taken to address the possible effect on the differential light curves, of any variation in the seeing disc during the session, since that might lead to spurious claims of INOV from such AGN due to the possibility of a significant contribution from the host galaxy to the total optical emission. From our observations, a duty cycle (DC) of INOV detection in the $\gamma$-ray NLSy1s  is estimated to be around 25\% - 30\%, which is comparable to that known for blazars. This estimate of DC will probably need an upward revision, once it becomes possible to correct for the dilution of the AGN's nonthermal optical emission by the (much steadier) optical emission contributed not only by the host galaxy but also the nuclear accretion disc in these high Eddington rate accretors. Finally, we also draw attention to the possibility that sharp optical flux changes on sub-hour time scale are less rare for $\gamma$-ray NLSy1s, in comparison to blazars.

\end{abstract}



 \section{Introduction}

 \label{sec1.0}

The diversity of intensity variations across the electromagnetic spectrum is a prominent characteristic of active galactic nuclei (AGNs) and it has been extensively leveraged to probe their emission mechanisms occurring on physical scales that are currently inaccessible to direct imaging by any available technique~\citep[e.g.,][]{Urry1995PASP..107..803U, Wagner1995ARA&A..33..163W, Ulrich1997ARA&A..35..445U, Zensus1997ARA&A..35..607Z}. AGN variability on minutes to hour-like time scales in the optical waveband is termed as intra-night optical variability~\citep[INOV,][]{Gopal-Krishna1993A&A...271...89G}. Such variations have proved particularly useful for probing and characterising the jet activity in blazars and radio-loud narrow-line Seyfert 1 (NLSy1) galaxies, as well as for identifying any blazar-like signatures in radio-quiet AGNs~\citep[e.g.,][]{Miller1989Natur.337..627M, Gopal-Krishna1993A&A...271...89G, Gopal-Krishna1995MNRAS.274..701G, Jang1995ApJ...452..582J, Heidt1996A&A...305...42H, Bai1999A&AS..136..455B, Romero1999A&AS..135..477R, Fan2001A&A...369..758F, Stalin2004MNRAS.350..175S, Gupta2005A&A...440..855G, Carini2007AJ....133..303C, 2009AJ....138..991R, Goyal2012A&A...544A..37G, Goyal2013MNRAS.435.1300G, Kumar2017MNRAS.471..606K, Paliya2019JApA...40...39P, Ojha2019MNRAS.483.3036O}. In the case of blazars, INOV is generally associated with disturbances within their relativistic jets, e.g., the well-known  ``shock-in-jet'' model ~\citep[e.g., see][]{Marscher1985ApJ...298..114M, Hughes1991ApJ...374...57H, Qian1991A&A...241...15Q, Wagner1995ARA&A..33..163W, Marscher1996ASPC..110..248M}. However, helicity, precession or other geometrical effects associated with the jets~\citep[e.g., see][]{Camenzind1992A&A...255...59C, Gopal-Krishna1992A&A...259..109G}, perhaps linked to short-term instabilities on the surface of the accretion disc but propagated into the relativistic effects jets, have also been considered as  possible mechanisms for INOV in blazars~(\citealp[e.g., see][]{Chakrabarti1993ApJ...411..602C};~\citealp[]{Mangalam1993ApJ...406..420M};~\citealp[also,][]{Czerny2008MNRAS.386.1557C}).\par
As recently summarised in~\citet[][]{Gopal-Krishna2018BSRSL..87..281G}, prominent among the radio-quiet AGNs covered extensively in INOV studies are radio-quiet Seyfert galaxies~\citep{Carini2003AJ....125.1811C}, radio-quiet quasars~\citep{Gopal-Krishna2003ApJ...586L..25G}, and weak emission line quasars~\citep{Gopal-Krishna2013MNRAS.430.1302G, Chand2014MNRAS.441..726C, Kumar2015MNRAS.448.1463K, Kumar2016MNRAS.461..666K, Kumar2017MNRAS.471..606K}. A possible cause for the low-level INOV detections in some such cases is the transient formation of micro-arcsec scale, probably in poorly aligned optical synchrotron jets~\citep[see,][]{Gopal-Krishna2003ApJ...586L..25G, Stalin2004MNRAS.350..175S}. The possibility that misaligned relativistic jets could be hidden among radio-quiet AGNs has also been noted  in the context of NLSy1 galaxies~\citep[e.g.,][]{Foschini2011nlsg.confE..24F, Berton2018A&A...614A..87B}. Nonetheless, INOV studies of these (low-luminosity) AGNs continue to be sparse and have been limited to a small data set ~\citep{Miller2000NewAR..44..539M, Ferrara2001ASPC..224..319F, Klimek2004ApJ...609...69K, Liu2010ApJ...715L.113L, Paliya2013MNRAS.428.2450P, Kshama2017MNRAS.466.2679K, Ojha2018BSRSL..87..387O, Paliya2019JApA...40...39P, Ojha2020MNRAS.493.3642O}, even though they were discovered over 3 decades ago~\citep{Osterbrock1985ApJ...297..166O} and basically defined by the rather small width of the optical Balmer emission lines, with full width at half-maximum (FWHM) being $<$ 2000 km s$^{-1}$ for H${\beta}$~\citep{Osterbrock1985ApJ...297..166O, Goodrich1989ApJ...342..908G}, stronger permitted Fe~{\sc ii}, and weak [O$_{III}$] emission lines such that the flux ratio of [O$_{III}]_{\lambda5007}/H\beta$ is $<$ 3~\citep{Shuder-Osterbrock1981ApJ...250...55S}. With some possible exceptions, they also show strong [Fe VII] and [Fe X] emission lines~(\citealp[see, however]{Pogge2011nlsg.confE...2P},~\citealp[]{Cracco2016MNRAS.462.1256C}). In addition to a soft X-ray spectrum~\citep{Boller1996A&A...305...53B, Wang1996A&A...309...81W,  Grupe1998A&A...330...25G}, they also known to display rapid X-ray (and sometimes even optical) flux variability~\citep[e.g.,][]{Leighly1999ApJS..125..297L, Komossa-Meerschweinchen2000A&A...354..411K, Miller2000NewAR..44..539M, Klimek2004ApJ...609...69K, Liu2010ApJ...715L.113L, Paliya2013MNRAS.428.2450P, Kshama2017MNRAS.466.2679K, Ojha2019MNRAS.483.3036O}. Observational evidence, as reviewed recently in~\citet{Paliya2019JApA...40...39P}, suggests that this class of AGN harbour relatively less massive black holes~\citep[$10^{6} - 10^{8} M_{\sun}$, ][]{Grupe2004ApJ...606L..41G, Deo2006AJ....132..321D, Peterson2011nlsg.confE..32P} which are accreting at a high fraction of the Eddington rate, in contrast to quasars~\citep[e.g.,][]{Peterson2000ApJ...542..161P}. Interestingly, NLSy1 galaxies (NLSy1s) exhibit the radio-loud/radio-quiet bimodality, displayed by QSOs, and thus are radio-quiet in most cases~\citep[][and references]{Kellermann2016ApJ...831..168K}. Only about $\sim$ 7\% of NLSy1s are radio-loud with the radio-loudness parameter\footnote{Radio-loudness is usually parameterised by the ratio (R) of the rest-frame flux densities at 5 GHz and 4400\AA, being R$\leq$ 10 and $>$ 10 for radio-quiet and radio-loud quasars, respectively~\citep[e.g. see,][]{Stocke1992ApJ...396..487S, Visnovsky1992ApJ...391..560V, Kellermann1994AJ....108.1163K, Kellermann1989AJ.....98.1195K}.} R$_{5GHz} > $ 10~\citep{Komossa2006AJ....132..531K, Zhou2006ApJS..166..128Z, Rakshit2017ApJS..229...39R, Singh2018MNRAS.480.1796S}. For R$_{5GHz} >$ 100, the fraction drops further to just 2 - 3 per cent~(\citealp[see also]{Komossa2006AJ....132..531K},~\citealp[]{Zhou2002ChJAA...2..501Z};~\citealp[]{Yuan2008ApJ...685..801Y}).\par
It may be noted that the inference about NLSy1s having relatively less massive black holes (BH), although not uncontested~\citep[e.g.,][]{Decarli2008MNRAS.386L..15D, Marconi2008ApJ...678..693M, Calderone2013MNRAS.431..210C, Viswanath2019ApJ...881L..24V}, appears nonetheless to be favoured by the weight of evidence, as summarized in~\citet{Paliya2019JApA...40...39P}. A crucial implication then would be that the jet activity in NLSy1s is powered  by central engines that operate significantly differently from those in quasars and their blazar subset. It is therefore remarkable that the jetted NLSy1s (see below) and blazars engender very similar manifestations of their relativistic jets, such as superluminal motion~\citep[see, e.g.,][]{Lister2018rnls.confE..22L}, a double-humped spectral energy distribution~\citep[e.g.,][]{Abdo2009ApJ...707L.142A, Paliya2013ApJ...768...52P, Paliya2019JApA...40...39P}, and violent optical and infrared flux variability~\citep{Liu2010ApJ...715L.113L, Itoh2013ApJ...775L..26I, Maune2013ApJ...762..124M, Paliya2013MNRAS.428.2450P}. This makes it especially desirable to search for any contrasting observational properties between blazars and the jetted NLSy1s. In this study, we pursue this objective from the standpoint of rapid optical flux variability, i.e., intranight optical variability (INOV) which is now believed to be a key attribute associated with blazar type AGN~\citep[see, e.g.,][]{Goyal2013MNRAS.435.1300G, Gopal-Krishna2018BSRSL..87..281G, Gopal-Krishna2019BSRSL..88..132G}.\par
It is noteworthy that in the case of NLSy1 galaxies, radio emission from star formation process alone could sometimes lead to a radio-loud classification, even if a relativistic jet is not present~\citep{Ganci2019A&A...630A.110G}. Therefore, a conclusive evidence that NLSy1s are capable of ejecting relativistic jets emerged only from the rather unexpected discovery of $\gamma$-ray emission from a handful of NLSy1s, using the {\it Fermi}-Large Area Telescope ({\it Fermi}-LAT)\footnote{https://heasarc.gsfc.nasa.gov/docs/heasarc/missions/fermi.html}~\citep{Abdo2009ApJ...699..976A, Abdo2009ApJ...707..727A, Abdo2009ApJ...707L.142A, Foschini2010ASPC..427..243F, Foschini2011nlsg.confE..24F, D'Ammando2012MNRAS.426..317D, D'Ammando2015MNRAS.452..520D, Yao2015MNRAS.454L..16Y, Paliya2018ApJ...853L...2P, Yang2018MNRAS.477.5127Y, Yao2019MNRAS.487L..40Y}. Since then, the number of such NLSy1s has slowly increased to just about 20~\citep{Paliya2019JApA...40...39P}, confirming their rarity. Hereinafter, we shall refer to such AGN as $\gamma$-ray NLSy1.\par
In fact, almost a decade prior to their detection of NLSy1s in  $\gamma$-rays, INOV had already been detected from a few NLSy1s~\citep{Miller2000NewAR..44..539M, Ferrara2001ASPC..224..319F, Klimek2004ApJ...609...69K}. The first INOV detection in a {\it Fermi}/LAT detected NLSy1 galaxy was for J094857.30$+$002224.0, for which ~\citet{Liu2010ApJ...715L.113L} observed a huge INOV amplitude of $\sim$ 0.5-mag. This source is known to exhibit strong optical variability on longer time scales as well~\citep{Maune2013ApJ...762..124M}. Nonetheless, as an AGN class, $\gamma$-ray NLSy1s are marked by rather poorly known INOV characteristics, primarily because INOV observations have so far been reported for just 6\footnote{We exclude the NLSy1 galaxy J110223.37$+$223920.5, with $<$ 3$\sigma$ detection in $\gamma$-rays~\citep{Foschini2015A&A...575A..13F}, whose INOV observations are reported by~\citet{Ojha2020MNRAS.493.3642O}.} of them~\citep{Liu2010ApJ...715L.113L, Paliya2013MNRAS.428.2450P, Paliya2016ApJ...819..121P, Kshama2017MNRAS.466.2679K, Ojha2019MNRAS.483.3036O, Ojha2020MNRAS.493.3642O}. Therefore, the primary aim of the present study is to bridge this knowledge gap, by determining the INOV characteristics of this rare and enigmatic class of AGN. For this, we use a much larger sample, now possible to assemble, and employ a single telescope, as well as a uniform analysis procedure for the entire sample.\par
The paper is structured as follows. In Sect.~\ref{section_2.0}, we outline the sample selection procedure. Sect.~\ref{section_3.0} provides details of our intra-night optical monitoring and the  data reduction procedure. The statistical analysis is presented in Sect.~\ref{sec4.0} and our main results followed by a brief discussion are given in Sect.~\ref{sec 5.0}. In Sect.~\ref{sect 6.0}, we summarise our main conclusions.

 \begin{table}
  \begin{minipage}{85mm} 
 \begin{center}   

   \caption[caption]{The present sample of 15 $\gamma-$ray detected NLSy1 galaxies. Columns are as follows: (1) SDSS name of the sources; (2) R-band magnitude, taken from~\citet{Monet1998AAS...19312003M}; (3) emission-line redshift, taken from~\citet{Paliya2019ApJ...872..169P}, and (4) $R_{1.4 GHz}\equiv f_{1.4 GHz}/f_{4400\AA}$, estimated by taking the rest-frame 1.4 GHz flux density from~\citet{Paliya2019JApA...40...39P} and the optical flux density at rest-frame 4400\AA~, estimated by fitting the SDSS spectrum, following the procedure described in~\citet{Rakshit2017ApJS..229...39R}. Note that the values of $R_{1.4 GHz}$ marked with an `$\star$' are taken from~\citet{Foschini2011nlsg.confE..24F}, except for J164442.53$+$261913.3 for which the value is taken from~\citet{Yuan2008ApJ...685..801Y} and J122222.99$+$041315.9 for which $R_{1.4 GHz}$ has been estimated using its core flux density of 0.6 Jy at 1.4 GHz from~\citet{Yuan2008ApJ...685..801Y}.}

\label{tab:source_info}
\begin{tabular}{rlrr}
 \hline

\multicolumn{1}{c}{SDSS Name} & \multicolumn{1}{c}{R-mag} &  \multicolumn{1}{c}{$z$} & \multicolumn{1}{r}{$R_{1.4 GHz}$} \\ 
 \multicolumn{1}{c}{(1)}& \multicolumn{1}{c}{(2)} & \multicolumn{1}{c}{(3)} & \multicolumn{1}{c}{(4)}\\
\hline
 J032441.20$+$341045.0    & 13.10 & 0.06  &  318$^{\star}$    \\
 J084957.98$+$510829.0    & 17.79 & 0.58  & 4496$^{\star}$  \\
J093241.15$+$530633.8     & 18.82 & 0.60 &19188    \\
J093712.33$+$500852.1     & 18.88 & 0.28 & 2007  \\
J094635.07$+$101706.1     & 18.90 & 1.00 & 11731\\
 J094857.32$+$002225.6   & 18.17 & 0.58 &  846$^{\star}$  \\
J095820.90$+$322401.6    & 15.51 & 0.53 & 1786   \\
 J122222.99$+$041315.9  & 17.06 & 0.97 & 1534$^{\star}$  \\
J130522.75$+$511640.2   & 15.80 & 0.79 & 509\\
J144318.56$+$472556.7   & 17.70 & 0.70 & 1921\\
 J150506.48$+$032630.8 & 17.72 & 0.41 & 3364$^{\star}$  \\
J152039.70$+$421111.0   & 18.09 & 0.48 &42908   \\
 J164442.53$+$261913.3  & 16.60 & 0.14 &  447$^{\star}$  \\
J211817.40$+$001316.8   & 18.60 & 0.46 &12410    \\
J211852.90$-$073229.3   & 19.48 & 0.26 & 4664    \\
\hline

\end{tabular}
 \end{center}
 \end{minipage}
\end{table}

\section{The sample}
\label{section_2.0}

The present sample is a well-defined, large subset of the compilation of all 22 NLSy1s reported to have a $\gamma$-ray detection~\citep{Paliya2019JApA...40...39P}. Our first filter, namely a $\gamma$-ray detection threshold  of $> 3\sigma$, led to the rejection of the sources  J124634.65$+$023809.1 and J142106.00$+$385522.5~\citep[see,][]{Foschini2011nlsg.confE..24F, Paliya2018ApJ...853L...2P}. A third source, J200755.18$-$443444.3, got excluded because its low declination precludes monitoring from our observatory.  Out of the remaining 19 sources, another two, namely J110223.37$+$223920.5 and J164100.1$+$345453.0 were rejected since their reported $\gamma$-ray detections have not been confirmed by~\citet{Foschini2015A&A...575A..13F} and~\citet{Ciprini2018rnls.confE..20C}. Note that these two sources are also not contained in the {\it Fermi} Large Area Telescope Fourth Source Catalog~\citep{Abdollahi2020ApJS..247...33A}. The sixth source to be excluded is J003159.9$+$093618.0, since~\citet{Paliya2019JApA...40...39P} suggests it to be a probable counterpart of  the unidentified $\gamma$-ray source  3FGL J003159$+$093615. The last source to be excluded is 3C 286 (J133108.3$+$303032.0); unlike all other $\gamma$-ray detected NLSy1s, its radio spectrum is steep and it has been argued that its $\gamma$-ray emission may substantially originate outside the jet~\citep{Berton2018A&A...614A..87B}. The final sample of 15 $\gamma$-ray NLSy1s is listed in Table~\ref{tab:source_info}.

\begin{table*}
  \caption{The observational log and basic parameters of the comparison stars used for the sample of 15 $\gamma-$ray NLSy1 galaxies. Columns are listed as follows: (1) Target AGN and the comparison stars; (2) date(s) of monitoring; (3) right ascension; (4) declination; (5) SDSS $g$-band magnitude; (6) SDSS $r$-band magnitude; (7) SDSS `$g-r$' colours. The positions and apparent magnitudes of the sources and their comparison stars were taken from the SDSS DR14~\citep{Abolfathi2018ApJS..235...42A}. Due to the non-availability of the SDSS `$g-r$' colours for the source, J032441.20$+$341045.0, and its comparison stars (marked by `$\dagger$' in columns 5, 6 \& 7), `B-R' colours have been used from USNO-A2.0 catalog~\citep{Monet1998AAS...19312003M} with  B-magnitude as 14.50 (target AGN), 15.60 (S1), 16.20 (S2) and R-magnitude as 13.70 (target AGN), 14.40 (S1), 14.40 (S2).}

  \label{tab_gray_comp_star}
\begin{tabular}{ccc ccc c}\\
\hline
{Target AGN and} &   Date(s) of monitoring       &   {R.A.(J2000)} & {Dec.(J2000)}              & {\it g} & {\it r} & \hspace{0.1 cm} {\it g-r} \\
the comparison stars           &      &   (hh mm ss)       &($^\circ$ $^\prime$ $^{\prime\prime}$)   & (mag)   & (mag)   & (mag)     \\
{(1)}      & {(2)}        & {(3)}           & {(4)}                              & {(5)}   & {(6)}   & \hspace{0.1 cm} {(7)}     \\
\hline
\multicolumn{7}{l}{}\\
J032441.20$+$341045.0 & 2016 Nov. 22, 23; Dec. 02; 2017 Jan. 03, 04& 03 24 41.20  &$+$34 10 45.00 & -----$^{\dagger}$ & -----$^{\dagger}$ & ---$^{\dagger}$   \\
S1                    &                                            & 03 24 53.68  &$+$34 12 45.62 & -----$^{\dagger}$ & -----$^{\dagger}$ & ---$^{\dagger}$    \\
S2                    &                                            & 03 24 53.55  &$+$34 11 16.58 & -----$^{\dagger}$ & -----$^{\dagger}$ & ---$^{\dagger}$    \\
J084957.98$+$510829.0 & 2017 Dec. 13, 2019 April 08& 08 49 57.98  &$+$51 08 29.04 & 18.92 & 18.28& 0.64\\
S1                    &                           & 08 50 12.62  &$+$51 08 08.03 & 19.45 & 18.06& 1.39  \\
S2                    &                           & 08 50 03.07  &$+$51 09 12.23 & 17.82 & 17.09& 0.73  \\
J093241.15$+$530633.8 & 2019 Jan. 13, 2020 April 11& 09 32 41.15  &$+$53 06 33.79 & 18.90 & 18.84& 0.06  \\
S1                    & 2019 Jan. 13              & 09 32 17.61  &$+$53 01 48.66 & 19.73 & 18.21& 1.52  \\
S2                    & 2019 Jan. 13              & 09 32 41.51  &$+$53 06 14.10 & 18.72 & 17.50& 1.22  \\
S3                    & 2020 April 11             & 09 32 12.62  &$+$53 09 09.46 & 17.86 & 16.82& 1.04  \\
S4                    & 2020 April 11             & 09 31 53.14  &$+$53 01 30.54 & 16.88 & 16.21& 0.67  \\
J093712.33$+$500852.1 & 2019 March 23             & 09 37 12.33  &$+$50 08 52.14 & 19.53 & 18.79& 0.74  \\
S1                    &                           & 09 38 01.04  &$+$50 08 49.80 & 18.36 & 17.84& 0.52  \\
S2                    &                           & 09 36 34.62  &$+$50 09 56.10 & 18.45 & 17.07& 1.38  \\
J094635.07$+$101706.1 & 2019 Dec., 26, 29         & 09 46 35.07  &$+$10 17 06.13 & 19.51 & 19.21& 0.30  \\
S1                    & 2019 Dec., 26, 29         & 09 46 50.04  &$+$10 10 13.97 & 18.25 & 17.71& 0.54    \\
S2                    & 2019 Dec., 26             & 09 46 35.29  &$+$10 11 40.29 & 18.02 & 17.64& 0.38    \\
S3                    & 2019 Dec., 29             & 09 47 04.35  &$+$10 15 52.01 & 18.72 & 18.03& 0.69    \\
J094857.32$+$002225.6 & 2016 Dec. 02; 2017 Dec. 21& 09 48 57.32  &$+$00 22 25.56 & 18.59 & 18.43& 0.16    \\
S1                    &                           & 09 48 36.95  &$+$00 24 22.55 & 17.69 & 17.28& 0.41    \\
S2                    &                           & 09 48 37.47  &$+$00 20 37.02 & 17.79 & 16.70& 1.09    \\
J095820.90$+$322401.6 & 2019 Jan. 08; Feb. 16     & 09 58 20.90  &$+$32 24 01.60 & 16.01 & 16.00& 0.01     \\
S1                    &                           & 09 58 18.30  &$+$32 28 34.43 & 15.80 & 15.31& 0.49    \\
S2                    &                           & 09 58 35.20  &$+$32 28 19.27 & 15.50 & 15.07& 0.43    \\
J122222.99$+$041315.9 & 2017 Jan. 03, 04; Feb. 21, 22; March 04, 24  & 12 22 22.99  &$+$04 13 15.95 & 17.02 & 16.80& 0.22  \\
S1                    &                                              & 12 22 34.02  &$+$04 13 21.57 & 18.63 & 17.19& 1.44      \\
S2                    &                                              & 12 21 56.12  &$+$04 15 15.19 & 17.22 & 16.78& 0.44    \\
J130522.75$+$511640.2 & 2017 April 04; 2019 April 25                 & 13 05 22.74  &$+$51 16 40.26 & 17.29 & 17.10& 0.19  \\
S1                    & 2017 April 04                                & 13 06 16.16  &$+$51 19 03.67 & 16.96 & 15.92& 1.04      \\
S2                    & 2017 April 04; 2019 April 25                 & 13 05 57.57  &$+$51 11 00.97 & 16.35 & 15.26& 1.09      \\
S3                    & 2019 April 25                                & 13 05 44.25  &$+$51 07 35.85 & 17.88 & 16.42& 1.46      \\
J144318.56$+$472556.7 & 2018 March 11, 23                        & 14 43 18.56  &$+$47 25 56.74 & 18.14 & 18.17& \hspace{-0.25 cm}$-$0.03  \\
S1                    &                                              & 14 43 37.14  &$+$47 23 03.03 & 17.51 & 16.82& 0.69      \\
S2                    &                                              & 14 43 19.05  &$+$47 19 00.98 & 18.03 & 16.75& 1.28      \\
J150506.48$+$032630.8 & 2017 March 25; 2018 April 12& 15 05 06.48  &$+$03 26 30.84 & 18.64 & 18.22& 0.42    \\
S1                    &                                             & 15 05 32.05   &$+$03 28 36.13 & 18.13 & 17.64& 0.49    \\
S2                    &                                             & 15 05 14.52   &$+$03 24 56.17 & 17.51 & 17.14& 0.37    \\
J152039.70$+$421111.0 & 2019 May 05, 2020 March 22& 15 20 39.70  &$+$42 11 11.19 & 19.31 & 18.94& 0.37 \\
S1                    & 2019 May 05               & 15 20 42.81  &$+$42 07 28.32 & 19.43 & 18.03& 1.40    \\
S2                    & 2019 May 05               & 15 21 20.78  &$+$42 03 20.45 & 17.51 & 17.09& 0.42    \\
S3                    & 2020 March 22             & 15 21 11.63  &$+$42 11 45.71 & 17.89 & 17.08& 0.81    \\
S4                    & 2020 March 22             & 15 20 57.17  &$+$42 10 04.87 & 18.13 & 16.68& 1.45    \\
J164442.53$+$261913.3 & 2017 April 03; 2019 April 26 & 16 44 42.53  &$+$26 19 13.3  & 18.03 & 17.61& 0.42    \\
S1                    &                              & 16 45 20.03  &$+$26 20 54.55 & 16.56 & 15.89& 0.67    \\
S2                    &                              & 16 44 34.40  &$+$26 15 30.27 & 16.28 & 15.80& 0.48    \\
J211817.40$+$001316.8 & 2019 June 07, 10             & 21 18 17.40  &$+$00 13 16.76 & 18.71 & 18.45& 0.26  \\
S1                    & 2019 June 07, 10             & 21 18 36.48  &$+$00 08 35.40 & 18.48 & 17.78& 0.70    \\
S2                    & 2019 June 07                 & 21 18 03.72  &$+$00 06 45.79 & 17.99 & 17.41& 0.58    \\
S3                    & 2019 June 10                 & 21 18 10.06  &$+$00 15 30.61 & 17.99 & 17.63& 0.36    \\
J211852.90$-$073229.3 & 2018 Oct.  04; 2019 June 09  & 21 18 52.90  &$-$07 32 29.34 & 18.78 & 17.96& 0.82  \\
S1                    &                              & 21 18 25.68  &$-$07 31 20.72 & 17.83 & 17.28& 0.55   \\
S2                    &                              & 21 18 52.33  &$-$07 31 37.64 & 17.78 & 16.88& 0.90    \\

\hline

\end{tabular}
\end{table*}

\begin{table*}
 \centering
 \begin{minipage}{190mm}
 {\small
   \caption[caption]{Observational details and the inferred INOV status for the sample of 15 $\gamma-$ray NLSy1 galaxies (photometric aperture radius used = 2$\times$FWHM). The columns are as follows: (1) SDSS name of the NLSy1; (2) date(s) of the monitoring session(s). Dates inside parentheses are for the longest two sessions which we have used for estimating the INOV duty cycle; (3) duration of the monitoring session; (4) number of data points in the DLCs for the monitoring session; (5) median seeing (FWHM in arcsec) for the session; (6) F-values for the two DLCs (relative to the two comparison stars), based on the F$^{\eta}$-test; (7) INOV status inferred from the two DLCs, using the F$^{\eta}$-test, with V = variable (confidence level $\geq$ 99\%), PV = probable variable (confidence level between 95-99\%),  NV = non-variable (confidence level $<$ 95\%); (8) F-values for the `NLSy1-reference star' DLC, based on the F$_{enh}$-test; (9) INOV status inferred using the F$_{enh}$-test; (10) mean photometric error for the two DLCs of the target AGN relative to the two comparison stars; (11) mean amplitude of variability in the two DLCs; and (12) exposure time per frame in the DLC.}

 \label{NLSy1:tab_result}
 \begin{tabular}{ccc ccccr ccrc}\\
   \hline
   {NLSy1s} &{Date(s)} &  {T}  & {N}  & Median & { $F^{\eta}$-test} & {INOV } & { $F_{enh}$-test} & {INOV} &{$\sqrt { \langle \sigma^2_{i,err} \rangle}$} & $\overline\psi_{s1, s2}$ & Exposure time \\
   (SDSS name) & yyyy.mm.dd & (hr) & & FWHM  & {$F_1^{\eta}$},{$F_2^{\eta}$} & status & $F_{enh}$ & status  & (AGN-s) & (\%)& for each science\\
   &&&& (arcsec) &           &{99\%}& & {99\%}& & & frame (sec)\\
    {(1)}&{(2)} & {(3)} & {(4)} & {(5)} & {(6)} & {(7)} & {(8)} & {(9)} & {(10)} & {(11)} & {(12)}\\
   \hline
   
 J032441.20$+$341045.0 & (2016.11.22) & 4.42 & 56& 2.32& 14.34, 16.13 &  V,  V   & 17.39 &   V  & 0.003 &  5.38  & 240\\   
                       &  2016.11.23  & 4.27 & 54& 2.13& 12.01, 10.69 &  V,  V   & 05.20 &   V  & 0.003 &  4.07  & 240    \\   
                       & (2016.12.02) & 4.41 & 44& 2.60& 85.80, 88.73 &  V,  V   & 98.29 &   V  & 0.003 & 11.44  & 300    \\   
                       &  2017.01.03  & 3.00 & 39& 2.47& 08.14, 10.55 &  V,  V   & 03.68 &   V  & 0.003 &  4.02  & 240    \\   
                       &  2017.01.04  & 3.39 & 33& 2.45& 35.03, 35.56 &  V,  V   & 17.16 &   V  & 0.003 &  7.49  & 300    \\   
 J084957.98$+$510829.0 & (2017.12.13) & 4.42 & 24& 2.83& 00.42, 00.51 &  NV, NV  & 00.77 &   NV & 0.033 &    --  & 600    \\   
                       & (2019.04.08) & 3.04 & 13& 2.88& 00.66, 00.89 &  NV, NV  & 00.62 &   NV & 0.032 &    --  & 840    \\   
 J093241.15$+$530633.8 & (2019.01.13) & 3.32 & 17& 2.86& 00.80, 00.81 &  NV, NV  & 04.09 &   V  & 0.027 &  6.59  & 720     \\  
                       & (2020.04.11) & 3.39 & 10& 5.24& 07.94, 08.10 &   V,  V  & 07.94 &   V  & 0.046 & 44.64  &\hspace{-0.15 cm} 1080 \\  
 J093712.33$+$500852.1 & (2019.03.23) & 3.40 & 14& 2.54& 00.43, 00.33 &  NV, NV  & 01.46 &   NV & 0.025 &    --  & 900    \\   
 J094635.07$+$101706.1 & (2019.12.26) & 4.03 & 14& 4.02& 00.30, 00.34 &  NV, NV  & 02.72 &   NV & 0.045 &    --  &\hspace{-0.15 cm}1020 \\   
                       & (2019.12.29) & 3.56 & 13& 3.73& 00.54, 00.42 &  NV, NV  & 01.64 &   NV & 0.041 &    --  & 900    \\   
 J094857.32$+$002225.6 & (2016.12.02) & 4.15 & 17& 2.58& 01.71, 01.88 &  NV, NV  & 10.52 &   V  & 0.017 &  7.95  & 780    \\   
                       & (2017.12.21) & 5.19 & 33& 2.24& 13.95, 16.53 &  V , V   & 25.26 &   V  & 0.012 & 16.42  & 540    \\   
 J095820.90$+$322401.6 & (2019.01.08) & 3.22 & 27& 2.81& 02.80, 03.00 &  V , V   & 01.17 &   NV & 0.004 &  3.00  & 360    \\   
                       & (2019.02.16) & 4.05 & 24& 2.97& 21.15, 21.86 &  V , V   & 24.90 &   V  & 0.009 & 12.98  & 540     \\  
 J122222.99$+$041315.9 &  2017.01.03  & 3.52 & 17& 2.38& 00.62, 00.30 &  NV, NV  & 00.68 &   NV & 0.018 &    --  & 600    \\   
                       &  2017.01.04  & 3.14 & 16& 2.36& 00.32, 00.37 &  NV, NV  & 01.99 &   NV & 0.014 &    --  & 720    \\   
                       &  2017.02.21  & 4.44 & 41& 2.65& 00.74, 00.76 &  NV, NV  & 02.13 &   V  & 0.020 &  6.36  & 360    \\   
                       & (2017.02.22) & 5.50 & 50& 2.59& 03.98, 03.60 &  V , V   & 06.51 &   V  & 0.017 & 13.33  & 360    \\   
                       & (2017.03.04) & 4.93 & 39& 2.61& 00.72, 00.86 &  NV, NV  & 01.36 &   NV & 0.019 &    --  & 360    \\   
                       &  2017.03.24  & 3.94 & 39& 2.37& 00.93, 00.75 &  NV, NV  & 01.66 &   NV & 0.020 &    --  & 360    \\   
 J130522.75$+$511640.2 & (2017.04.04) & 3.79 & 23& 2.57& 00.66, 00.70 &  NV, NV  & 02.94 &   PV & 0.012 &    --  & 600    \\   
                       & (2019.04.25) & 3.11 & 22& 2.77& 01.42, 01.56 &  NV, NV  & 04.12 &   PV & 0.018 &    --  & 480    \\   
 J144318.56$+$472556.7 & (2018.03.11) & 3.23 & 19& 3.15& 00.56, 00.54 &  NV, NV  & 02.59 &   NV & 0.022 &    --  & 480    \\   
                       & (2018.03.23) & 3.08 & 23& 2.33& 00.36, 00.35 &  NV, NV  & 01.00 &   NV & 0.018 &    --  & 480    \\   
 J150506.48$+$032630.8 & (2017.03.25) & 5.21 & 41& 2.08& 00.60, 00.59 &  NV, NV  & 01.04 &   NV & 0.028 &    --  & 420    \\   
                       & (2018.04.12) & 3.05 & 19& 2.55& 00.67, 00.63 &  NV, NV  & 00.84 &   NV & 0.032 &    --  & 480    \\   
 J152039.70$+$421111.0 & (2019.05.04) & 3.02 & 11& 3.08& 00.56, 00.61 &  NV, NV  & 01.43 &   NV & 0.031 &    --  & 900    \\   
                       & (2020.03.22) & 3.00 & 09& 4.85& 01.00, 01.00 &  NV, NV  & 03.69 &   NV & 0.034 &    --  & 900    \\   
 J164442.53$+$261913.3 & (2017.04.03) & 4.37 & 37& 2.50& 01.44, 01.28 &  NV, NV  & 03.53 &   V  & 0.011 &  5.41  & 420    \\   
                       & (2019.04.26) & 3.22 & 24& 2.27& 03.06, 03.74 &  V , V   & 06.40 &   V  & 0.011 &  7.50  & 480    \\   
 J211817.40$+$001316.8 & (2019.06.07) & 3.31 & 10& 2.19& 03.00, 03.08 &  NV, NV  & 26.26 &   V  & 0.046 & 29.75  &\hspace{-0.15 cm} 1200 \\   
                       & (2019.06.10) & 3.17 & 09& 2.64& 16.38, 18.42 &  V ,  V  &144.97 &   V  & 0.032 & 35.57  &\hspace{-0.15 cm} 1200 \\   
 J211852.90$-$073229.3 & (2018.10.04) & 4.47 & 13& 2.81& 01.87, 01.26 &  NV, NV  & 03.47 &   NV & 0.014 &    --  & 900    \\   
                       & (2019.06.09) & 3.15 & 09& 2.30& 06.38, 06.55 &  V , V   & 12.31 &   V  & 0.019 & 17.85  &\hspace{-0.15 cm} 1200  \\
   \hline
  
    \end{tabular} 
 
 }              
 \end{minipage} 
    \end{table*}

\begin{table}
 
 {\small
   \caption{The DC and $\overline{\psi}$ of INOV, computed for the present sample of 15 $\gamma-$ray NLSy1 galaxies, based on the $F_{enh}$-test and  $F^{\eta}$-test.}
   
 \label{NLSy1:DC_result}
\centering
 \begin{tabular}{ccccc}
   \hline
    &  \multicolumn{2}{ l }{Based on the $F_{enh}$-test} & \multicolumn{2}{ l }{Based on the $F^{\eta}$-test} \\
     \hline
        {$\gamma-$ray NLSy1s}   &{$^{\star}$DC}   &{$^{\star}\overline{\psi}^{\dag}$} &   {$^{\star}$DC}   &  {$^{\star}\overline{\psi}^{\dag}$}    \\
                                &({\%})         &  ({\%})                  &  ({\%})  &     ({\%})             \\
  \hline
                                &  39           &   16                     &   30     &   17       \\  
       
  \hline
  
  \multicolumn{5}{l}{$^{\star}$Only using the 29 sessions, as explained in Sect.~\ref{sec 4.1}.}\\
  \multicolumn{5}{l}{$^{\dag}$This is the mean value for all the DLCs belonging to the type `V'.}\\

 \end{tabular}  
 }             

\end{table}

\section{Observations and Data Reduction}
\label{section_3.0}

\subsection{Photometric monitoring observations}

All 15 $\gamma-$ray NLSy1s in our sample were monitored in the broadband filter R, with the 1.3 metre (m) Devasthal Fast Optical Telescope (DFOT) located at Devasthal near Nainital~ \citep{Sagar2010ASInC...1..203S} and operated by the Aryabhatta Research Institute of Observational Sciences (ARIES), India. The DFOT is a Ritchey-Chretien (RC) telescope with a fast beam (f/4) and has a pointing accuracy better than 10 arcsec rms.  It is equipped with a 2k$\times$2k deep thermoelectrically cooled (to about $-85^{\circ}$C) Andor CCD camera, has a pixel size of 13.5 microns, a plate scale of 0.53 arcsec per pixel, giving a field-of-view (FOV) of $\sim$18$\times$18 arcmin$^{2}$ on the sky. The detector chip, having a system rms noise and gain of 7.5 $e^-$ and 2.0 $e^-$ ADU$^{-1}$, respectively, was read out at the speed of 1 MHz. Each target AGN was monitored for a duration between 3.0 to 5.5 hours (hrs), as is usual for our INOV campaigns, with the aim to enhance the probability of INOV detection~\citep[see][]{Carini1990PhDT.......263C}. Also, in order to strengthen the INOV statistics, each of our 15  $\gamma-$ray NLSy1s was monitored at least in two intranight sessions, the solitary  exception being the source J093712.33$+$500852.1 for which bad weather restricted the monitoring to a single successful session. The exposure time for each science frame was set between 4 and 20 minutes, depending on the brightness of the AGN, the sky transparency, and the lunar phase. The typical median seeing (FWHM of the point spread function (PSF)) during our monitoring sessions ranged between $\sim$ 2 - 3 arcsec.

\subsection{Data reduction}
\label{sec3.2}

Pre-processing of the raw frames was done by performing the bias subtraction, flat-fielding, and cosmic-ray removal using the standard tasks within the {\textsc IRAF}\footnote{Image Reduction and Analysis Facility (http://iraf.noao.edu/)} software package.  Instrumental magnitudes of the target NLSy1 and the two chosen comparison stars registered in the CCD frames were measured by aperture photometry~\citep{1987PASP...99..191S, 1992ASPC...25..297S}, using the DAOPHOT II algorithm\footnote{Dominion Astrophysical Observatory Photometry (http://www.astro.wisc.edu/sirtf/daophot2.pdf)}. A key parameter for photometry is the aperture size, used for measuring the instrumental magnitude and the corresponding signal-to-noise ratio (S/N) of the individual photometric data points. As reported in~\citet{Howell1989PASP..101..616H}, S/N is maximised  when the photometric aperture radius is approximately equal to the FWHM of the seeing disc, i.e., Point-Spread-Function (PSF). However, we note that the situation can be more complex while dealing with some relatively nearby AGNs (e.g., two sources in our sample, viz., J032441.20$+$341045.0 at z=0.06; J164442.53$+$261913.3 at z=0.14), because a significant contribution to the total flux can come from the underlying host galaxy and hence the relative contributions of the (point-like) AGN and the host galaxy to the aperture photometry can vary significantly as the PSF changes during the session. As emphasised by~\citet{Cellone2000AJ....119.1534C}, the standard analysis of the differential light curves (DLCs) could then lead to statistically significant, yet spurious claims of INOV for small apertures comparable to the PSF.  Keeping this caution in mind, we have first estimated the PSF (FWHM) for each CCD frame (by averaging the profiles of 5 bright although clearly unsaturated stars in that frame) and treating the median of those values as the PSF (i.e., FWHM) for that session. The photometry was then performed taking four values of aperture radius, viz., 1$\times$FWHM, 2$\times$FWHM, 3$\times$FWHM and 4$\times$FWHM. The resulting DLCs of the target AGN were compared with the observed trend of variation of the seeing disc (FWHM) during the session, before proceeding with a quantitative analysis. In practice, this turned out to be an extra cautious approach since, except for three sessions (dated: 2019.04.25, 2019.05.04 and 2020.03.22 see Figs.~\ref{fig:lurve 3},~\ref{fig:lurve 4}), FWHM remained quite steady, (variations mostly within an arcsec), without showing any systematic trend during the session. Under this circumstance, an aperture radius of 2$\times$FWHM is expected to yield fairly reliable DLCs for the target AGN, even if its host galaxy is up to 2-mag brighter than the AGN~\citep[see, table 2 of][]{Cellone2000AJ....119.1534C}.\par
As an additional check for a significant host-galaxy contribution to our aperture photometry, we have fitted the brightness profile of a bright (unsaturated) star, placed at the position of our target AGN. The brightness of the star was varied while keeping the FWHM unchanged. The latter was determined by taking the median of the values measured for nearby 10 bright (unsaturated) stars within the same CCD frame. The best-fit profile obtained was then subtracted out from the target image, to get the residual image. The counts in the residual image were found to be comparable to the sky background counts for all our sources, except for J032441.20$+$341045.0 with the highest significance reaching 2.13$\sigma$ for the monitoring session dated 2016-11-23. Note that this AGN is a low-redshift NLSy1 galaxy ($z = 0.06$), and is already known to have a clearly visible host galaxy, as discussed in details by~\citet{Ojha2019MNRAS.483.3036O}. This basic trend emerging from the model-fitting is in accord with the findings of the recent search for host galaxy emission for NLSy1 galaxies~\citep{2020MNRAS.492.1450O}. We note that 9 of our NLSy1s are in fact common to their sample and they could detect host galaxies of just 4 of them, which are also the nearest ($z \lesssim$ 0.6) ones in their sample. And this is in spite of their using a far more sensitive telescope (the 8.6-m ESO Very Large Telescope (VLT)) than ours and taking much longer exposures (median 1800 seconds) than we took (median 540 seconds) in our observations for our current sample with the 1.3-m DFOT. Additional comments on some individual cases including this low redshift source are provided in Sect.~\ref{sec 5.0}.\par
For each session, DLCs of the target NLSy1 were derived relative to two (steady) comparison stars which were chosen on the basis of their proximity to the target AGN in brightness, the importance of which for a reliable INOV detection has been highlighted by~\citet{Howell1988AJ.....95..247H} and further emphasised in~\citet{Cellone2007MNRAS.374..357C}. For 8 sources of our sample, we were able to find at least one comparison star which falls within $\sim$ 1-mag of the target AGN. The magnitude offsets ($\Delta m_{R}$) for the remaining sources are also not large, namely 2.0-mag (J093241.15$+$530633.8), 1.57-mag (J094635.07$+$101706.1), 1.26-mag (J130522.75$+$511640.2), 1.29-mag (J144318.60$+$472557.0), 1.53-mag (J152039.70$+$421111.00), 1.54-mag (J164442.53$+$261913.3 and 1.40-mag (J211817.40$+$001316.8). The coordinates and other parameters of the comparison stars used for the different sessions are given in Table~\ref{tab_gray_comp_star} for our entire set of 15 $\gamma-$ray NLSy1s. The SDSS {\it $g-r$} colours for the target NLSy1 and their chosen comparison stars are:  $<$ 0.82 and $<$ 1.80 in all cases, with the median values being 0.30 and 0.70, respectively (column 7 in Table~\ref{tab_gray_comp_star}). It has been demonstrated by~\citet{Carini1992AJ....104...15C} and~\citet{Stalin2004MNRAS.350..175S} that colour differences of this order should produce a negligible effect on the DLCs as the  atmospheric attenuation varies during a session.

\section{STATISTICAL ANALYSIS}
\label{sec4.0}
To search for the presence/absence of INOV in a DLC, we have employed two different forms of the \emph{F$-$test} proposed by~\citet{Diego2010AJ....139.1269D}. These are: (i) the standard \emph{F$-$test}~\citep[hereafter $F^{\eta}$-test, e.g., see][]{Goyal2012A&A...544A..37G} and (ii) the power-enhanced \emph{$F-$test}~\citep[hereafter $F_{enh}$-test, e.g., see][]{Diego2014AJ....148...93D}. Detailed description of these two tests is given in our previous papers~\citep[][and references therein]{Ojha2019MNRAS.483.3036O, Ojha2020MNRAS.493.3642O}.\par

In brief, following~\citet{Goyal2012A&A...544A..37G}, $F^{\eta}$-test can be written as 

\begin{equation} 
 \label{eq.fetest}
\hspace{1.5cm} F_{1}^{\eta} = \frac{\sigma^{2}_{(q-s1)}} { \eta^2 \langle
  \sigma_{q-s1}^2 \rangle}, \nonumber \\
\hspace{0.1cm} F_{2}^{\eta} = \frac{\sigma^{2}_{(q-s2)}} { \eta^2 \langle
  \sigma_{q-s2}^2 \rangle}  \nonumber \\ 
\end{equation}

where $\sigma^{2}_{(q-s1)}$ and $\sigma^{2}_{(q-s2)}$ are the variances with $\langle \sigma_{q-s1}^2 \rangle=\sum_ {i=1}^{N}\sigma^2_{i,~err}(q-s1)/N$ and $\langle \sigma_{q-s2}^2 \rangle$ being the mean square (formal) rms errors of the individual data points in the `target AGN - comparison star1' and `target AGN - comparison star2' DLCs, respectively. The parameter $\eta$ is the error scaling factor and is taken to be $1.5$ from~\citet{Goyal2012A&A...544A..37G}~\citep[see][]{Ojha2020MNRAS.493.3642O}. The DLCs of the target AGN relative to the two comparison stars and the `star-star' DLCs are displayed in the $2^{nd}$, $3^{rd}$ and $4^{th}$ panels from the bottom, in Figs.~1-4, respectively.\par
For the current study, we have set two critical significance levels, $\alpha= 0.01$ and $ \alpha = 0.05$ which correspond to the confidence levels of 99\% and 95\%, respectively. The $F$-values, obtained for individual DLCs using Eq.~\ref{eq.fetest} were compared with the adopted critical $F$-value ($F_{c}$), and an NLSy1 is deemed to be variable only if the $F$-values computed for both its DLCs are above the critical $F$ value at 99\% confidence level (hereafter, $F_{c}(0.99)$). The computed $F^{\eta}$-values and the correspondingly inferred variability status for the 36 sessions devoted to the 15 $\gamma-$ray NLSy1s  are given in columns 6 and 7 of Table~\ref{NLSy1:tab_result}.\par   

The $F_{enh}$-test, as described in~\citet{Ojha2020MNRAS.493.3642O} can be written as

\begin{equation}
\label{Fenh_eq}  
\hspace{1.0cm} F_{{\rm enh}} = \frac{s_{{\rm qso}}^2}{s_{\rm stc}^2}, \hspace{0.5cm} s_{\rm stc}^2=\frac{1}{(\sum _{j=1}^k N_j) - k}\sum _{j=1}^{k}\sum _{i=1}^{N_j}s_{j,i}^2
\end{equation}

where $s_{{\rm qso}}^2$ is  the variance of the DLC of the target AGN and the reference star (the one matching better in magnitude to the target AGN out of the two chosen comparison stars), while $s_{\rm stc}^2$ is the  stacked variance of the DLCs of the comparison stars and the reference star~\citep{Diego2014AJ....148...93D}. $N_{j}$ is the number of observations of the $j^{th}$ star and $k$ is the total number of comparison stars.\par

The $s_{{\rm j,i}}^2$ is the scaled square deviation defined as

\begin{equation}
\hspace{2.7cm} s_{j,i}^2=\omega _j(m_{j,i}-\bar{m}_{j})^2
\end{equation}

where $m_{j,i}$'s are the differential instrumental magnitudes, and $\bar{m_{j}}$ is the mean differential magnitude of the reference star and the j$^{th}$ comparison star. The scaling factor $\omega_ {j}$~\citep[see also][]{Joshi2011MNRAS.412.2717J} is taken as

\begin{equation}
 \hspace{2.7cm} \omega _j=\frac{\langle\sigma^2_{i,err}(q-ref)\rangle}{\langle\sigma^2_{i,err}(s_{j}-ref)\rangle}.
  \end{equation}

The $F_{{\rm enh}}$ value is estimated using Eq.~\ref{Fenh_eq} and compared with its $F_{c}(0.99)$ and $F_{c}(0.95)$. An NLSy1 is considered variable (V) and probable variable (PV) if the computed values of `NLSy1-reference star' DLC are found to be as $F_{{\rm enh}} > F_{c}(0.99)$ and $F_{c}(0.95) < F_{{\rm enh}} \leq F_{c}(0.99)$, respectively. The estimated $F_{{\rm enh}}$-values and the corresponding variability status for the 15 $\gamma-$ray NLSy1s are given in columns 8 and 9 of Table~\ref{NLSy1:tab_result}.

\subsection{Estimation of the INOV duty cycle and amplitude}
\label{sec 4.1}

For computing the duty cycle (DC) of INOV in our sample of 15 $\gamma-$ray NLSy1s, we have followed the definition given by~\citet{Romero1999A&AS..135..477R}

\begin{equation} 
\hspace{2.5cm} DC = 100\frac{\sum_{j=1}^n F_j(1/\Delta t_j)}{\sum_{j=1}^n (1/\Delta t_j)} 
\hspace{0.1cm}{\rm per~cent} 
\label{eqno1} 
\end{equation} 

where $\Delta t_j = \Delta t_{j,~obs}(1+$z$)^{-1}$ is the actual duration of the $j^{th}$ monitoring session, corrected for the target AGN's redshift, $z$~\citep[see details in][]{Ojha2019MNRAS.483.3036O, Ojha2020MNRAS.493.3642O}. $F_j$ is set equal to 1 if INOV is detected, otherwise, $F_j$ is taken to be zero. Secondly, in order to prevent the DC estimate from getting biased towards more frequently monitored sources in the sample, we have chosen to select just two sessions per source for calculating DC, even though data are available for more sessions (Table~\ref{NLSy1:tab_result}). The selection of the two sessions was kept unbiased, by basing it on the monitoring duration (T) and thus the longest two sessions  were selected, as shown by placing the session dates within parentheses on the panels in Figs.~1-4 and also in Table~\ref{NLSy1:tab_result}. The computed values for the sample of 15 $\gamma-$ray NLSy1, using the two different statistical tests are tabulated in Table~\ref{NLSy1:DC_result}.\par

For estimating the amplitude of INOV ($\psi$) of the monitored AGN, which quantifies its variation in a given session, we followed the definition given by~\citet{Heidt1996A&A...305...42H}

\begin{equation} 
\hspace{2.5cm} \psi= \sqrt{({A_{max}}-{A_{min}})^2-2\sigma^2}
\end{equation} 

with $A_{min,~max}$ = minimum (maximum) values in the `target AGN - star` DLC and $\sigma^2 = \eta^2\langle\sigma^2_{q-s}\rangle$, where, $\langle\sigma^2_{q-s}\rangle$ is the mean square (formal) rms errors of individual data points and $\eta$=1.5~\citep{Goyal2012A&A...544A..37G}.

\section{Results and discussion}
\label{sec 5.0}

As mentioned in Sect.~\ref{sec1.0}, the present INOV study based on a sample of 15 $\gamma-$ray NLSy1 galaxies represents a substantial advance over the previous reports of INOV properties of this rather poorly understood class of AGN. Those earlier investigations were based on just a handful (6)  of AGNs in this class. Recall that just two out of all the NLSy1s currently known to have a confirmed ($> 3\sigma$) detection in $\gamma$-rays have been left out of the present sample, on grounds well-justified in Sect.~\ref{section_2.0}. Thus, we believe that the INOV characterisation presented here should be fairly representative of this intriguing class of jetted AGN. This characterisation would also enable their systematic comparison with the already well established INOV properties of the blazar type AGN whose relativistic jets are not only more powerful, on the whole,  but also thought to form under qualitatively different operating conditions of the central engine (Sect.~\ref{sec1.0}). Another potential difference between these two AGN classes rests on the premise that the jetted NLSy1s represent an earlier stage in the evolution of flat-spectrum quasars/blazars~\citep[e.g.,][]{Mathur2000MNRAS.314L..17M, Sulentic2000ApJ...536L...5S, Mathur2001NewA....6..321M, Fraix-Burnet2017FrASS...4....1F, Komossa2018rnls.confE..15K, Paliya2019JApA...40...39P}.\par
\begin{figure*}
\centering
\epsfig{figure=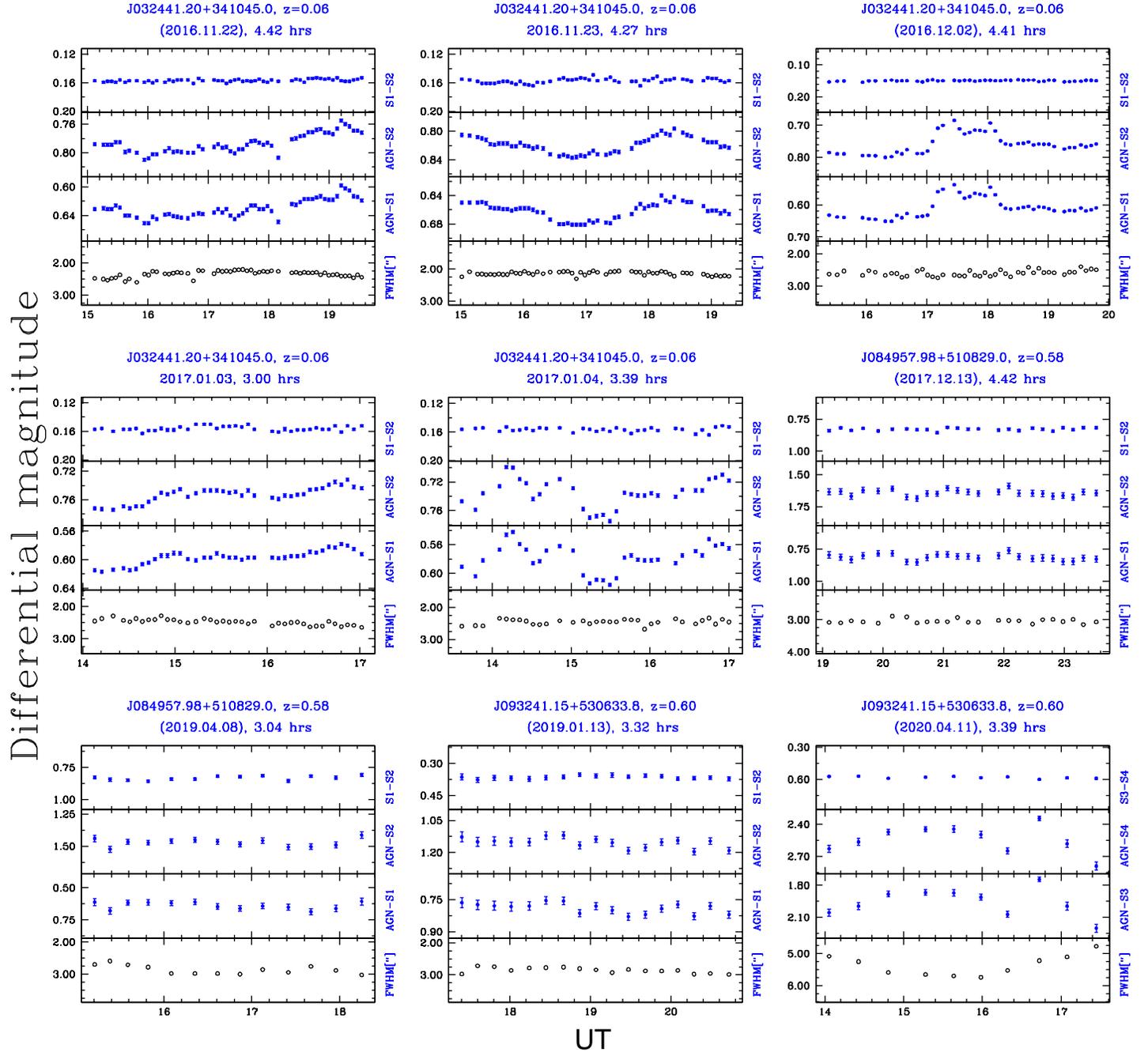}
\caption[]{Intranight differential light curves (DLCs) of the  first 3 $\gamma-$ray NLSy1s from our sample of 15 $\gamma-$ray NLSy1s. The AGN name, its redshift, and some observational details are given at the top of each panel. The sessions with the date given inside parentheses at the top of each panel were taken for the statistical analysis. In each panel, the upper DLC is derived using the chosen two (non-varying) comparison stars, while the lower two DLCs are the `NLSy1-star' DLCs, as defined  in the labels on the right side. The bottom panel displays the variations of the seeing disc (FWHM) during the monitoring session. }
  
\label{fig:lurve 1}
\end{figure*}

Out of the total 36 sessions, INOV was significantly detected by both F$^{\eta}$-test and the enhanced F-test ($F_{enh}$-test) in 12 sessions. For another 5 sessions, the  claim of INOV detection rests on the $F_{enh}$-test alone, while for one session only on the F$^{\eta}$-test.  Being mindful of this, we shall base our present discussion on the (more conservative) F$^{\eta}$-test. Another reason for this stance is that the available INOV information for large sets of blazars, too, is based on this test (see below).  But, before proceeding further, let us revisit the issue, already touched upon in Sect.~\ref{sec3.2}, that the optical aperture photometry of the AGN in some of low-z NLSy1s may have been significantly affected by a (varying) contribution from the host galaxy, in case the seeing disc (PSF) varied during the session. As emphasised by~\citet{Cellone2000AJ....119.1534C}, under such condition some DLCs may show statistically significant INOV which is actually spurious (see Sect.~\ref{sec3.2}). As seen from  Table~\ref{NLSy1:tab_result} and Figs.~1-4, data for 29 out of the total 36 sessions have actually been used here for INOV characterisation (see Sect.~\ref{sec 4.1}). Statistically significant INOV was detected in 13 of the 29 sessions, based on the  $F_{enh}$-test and in 10 sessions using the F$^{\eta}$-test. Since only the 10 sessions are used here for INOV statistics, we shall presently focus on their DLCs alone, from the viewpoint of a possible impact of any PSF variations. From Figs.~1-4,  it is seen  that non-negligible systematic PSF variation occurred in just 4 out of the 10 sessions. Their dates and the target AGNs are: 2020.04.11 (J093241.15$+$530633.8), 2017.02.22, (J122222.99$+$041315.9), 2019.06.09 (J211852.90$-$073229.3) and 2019.06.10 (J211817.40$+$001316.8). A closer inspection of the DLCs of these 4 sessions shows that, except for the one dated 2017.02.22, PSF actually remained fairly steady over the time span when AGN flux showed variation (Figs.~\ref{fig:lurve 1},~\ref{fig:lurve 4} ).  Only during the session on 2017-02-22 is the gradient of the DLCs of the target AGN (J122222.99$+$041315.9) seen to overlap in time with the systematic trend found in the PSF. However, the two gradients are seen to anti-correlate (Fig.~\ref{fig:lurve 3}), which is opposite to what is expected in case the aperture photometric measurements were getting significantly contaminated by the underlying galaxy~\citep[see,][]{Cellone2000AJ....119.1534C}. Even otherwise, any contamination from the host galaxy can not be significant for this AGN,  since its large redshift ($z$ = 0.97, Table~\ref{tab:source_info}) implies a high intrinsic optical luminosity, enough to swamp the host galaxy. Thus, to sum up, it can be asserted that the inferred INOV detection in neither of the 10 sessions is spurious, and this lends credence to the statistical properties derived using the present dataset. This claim is also in accord with the recent  high-quality imaging of NLSy1 galaxies by~\citet{2020MNRAS.492.1450O}, as noted in Sect.~\ref{sec3.2}.\par
The computed INOV DC and amplitudes ($\overline{\psi}$), using our 29 monitoring sessions devoted to the 15 $\gamma$-ray NLSy1s, are listed in Table~\ref{NLSy1:DC_result}. To recall, these are based on the application of the $F_{enh}$ and the $F^{\eta}$-tests, adopting a 99\% confidence level threshold for confirming INOV detection  (Sect.~\ref{sec4.0}). The two tests have yielded INOV duty cycles of 39\% and 30\%, respectively, for our sample. The corresponding estimates using a photometric aperture radius of 3xFWHM (instead of 2xFWHM), for which the DLCs are noisier, are found to be 35\% and 20\%, respectively. Thus, it is realistic to infer that the INOV DC for  $\gamma$-ray NLSy1s is at least around 25\%, based on the F$^{\eta}$-test and around 35\% to 40\% using the $F_{enh}$-test. In a systematic study of INOV characteristics of several prominent classes of powerful AGN, based on the $F^{\eta}$-test with a confidence level threshold also set at 99\%, ~\citet{Goyal2013MNRAS.435.1300G} showed that an INOV DC in excess of 20\% is observed exclusively for blazar type objects, i.e., high-optical-polarization quasars, BL Lacs, and TeV detected blazars~\citep[see, also][]{Gopal-Krishna2011MNRAS.416..101G}, their DC values being about 38\%, 40\% and 47\%, respectively. We may recall that, like the present campaign, the monitoring observations leading to these estimates were also made with 1-2 metre class telescopes and a detection threshold of the order of a few per cent was typically achieved. Thus, it seems that, as an AGN class,  $\gamma$-ray NLSy1s are well matched to blazars in terms of the occurrence rate of INOV, even though their central engines operate in a regime of distinctly higher Eddington accretion rate, as inferred widely in the literature (see Sect.~\ref{sec1.0}). One interesting consequence of the higher accretion rate for  $\gamma$-ray NLSy1s is the expected enhancement of the thermal component of the AGN's optical  emission, relative to its synchrotron emission~\citep[e.g., see][]{Zhou2007ApJ...658L..13Z, Paliya2014ApJ...789..143P}. Since the former is likely to be much less variable than the optical flux contributed by the (Doppler boosted) synchrotron jet, the fractional intranight optical variability (i.e., INOV amplitude $\psi$) would be suppressed due to the thermal contamination. The suppression can get compounded due to yet another contaminant,  namely the steady optical emission coming from the host galaxy (if detected), which may be significant in the case of low redshift (z $\lesssim$ 0.3) $\gamma$-ray NLSy1s harbouring intrinsically much weaker jets than blazars. Some likely implications of both these contaminating/diluting processes have been quantitatively examined recently by~\citet{Ojha2019MNRAS.483.3036O}, for the case of the prominent nearby $\gamma$-ray NLSy1 galaxy J032441.20$+$341045.0. The resulting prognosis in the present context is that future studies of $\gamma$-ray NLSy1s may well reveal an even stronger INOV, with a higher duty cycle, once it becomes possible to subtract out the contaminating thermal optical flux originating from the host galaxy and from the processes related to the AGN's inner accretion disc. Thus, it appears that the INOV activity in $\gamma$-ray NLSy1s may well turn out to be truly striking, even by the blazar standards.\par

We now turn attention to any sharp features present in the differential light curves, that are linked to the relativistic jet. In various intranight monitoring campaigns covering dozens of blazars, investigators have searched for brightness changes on time scales substantially shorter than an hour~\citep[e.g.,][]{Gopal-Krishna2011MNRAS.416..101G}. Such ultra-short time scales are important since the implied physical size closely approaches the event horizon of a $10^{8} - 10^{9} M_{\sun}$ black hole (\citealp[see, also]{Wiita2006ASPC..350..183W},~\citealp{Armitage2003MNRAS.341.1041A}). Indeed, flux variability on minute-like time scales has been convincingly detected in $\gamma$-ray emission from some blazars, e.g., PKS 2155$-$304~\citep{Aharonian2007ApJ...664L..71A} and PKS 1222$+$216~\citep{2011ApJ...730L...8A}. However, several hundred intranight optical monitoring sessions targeting blazars have yielded just a few claims of INOV on the minute-like time scale and it is clear that such events are an extremely rare occurrence in blazar light curves~\citep{Gopal-Krishna2018BSRSL..87..281G, Gopal-Krishna2019BSRSL..88..132G}. A remarkable case of sharp optical variation, with an implied  {\it  flux doubling time} of just $\sim$ 1 hr was observed in our monitoring of the $\gamma$-ray NLSy1 galaxy J032441.20$+$341045.0 on 2016-12-02 (see Fig.~\ref{fig:lurve 1} for the DLCs). There we had argued that after allowing for the actual dilution by the thermal optical emission, the flux doubling time could be even shorter by a substantial margin, perhaps approaching the minute-like time scales~\citep{Ojha2019MNRAS.483.3036O} observed in the $\gamma$-ray light curves of some blazars (see, above). In that study, we also pointed out that a similarly ultra-short flux doubling time can probably also be inferred from this AGN's already existing intranight DLCs dated 2012-12-09, published in~\citet{Paliya2014ApJ...789..143P} and, secondly, that such extremely rapid events may be correlated with the radio jet's superluminal speed~\citep{Ojha2019MNRAS.483.3036O}. Here we note that radio jets showing superluminal speeds have been shown to exist in 5 members of the present sample of  15 $\gamma$-ray NLSy1s. These five sources are: J032441.20$+$341045.0, J122222.99$+$041315.9, J150506.48$+$032630.8, with $v_{app}/c$ of $9.1\pm0.3$, $0.9\pm0.3$, $1.1\pm0.4$, respectively~\citep[e.g., see][]{Lister2016AJ....152...12L}, and  J084957.98$+$510829.0, J094857.32$+$002225.6 having  $v_{app}/c$ of $6.6\pm0.8$, $9.1\pm0.3$, respectively~\citep[e.g., see][]{Lister2019ApJ...874...43L}.  Optical flux variability on minute to hour-like time scales has also been reported for these 5 $\gamma$-ray NLSy1s~\citep{Liu2010ApJ...715L.113L, Paliya2013MNRAS.428.2450P, Paliya2016ApJ...819..121P, Ojha2019MNRAS.483.3036O}. The possibility of such a correlation could be firmly tested when the more extensive INOV database and superluminal speeds of the jets become available for $\gamma$-ray NLSy1s. In the present campaign,  we have searched for additional clean examples of variability on time scales $\ll$ 1 hr (see Figs. 1-4).  As explained below, only two such events could be identified, although some more may well be found when DLCs with comparably dense sampling become available for our entire sample .\par
\begin{figure*}
\centering
\epsfig{figure=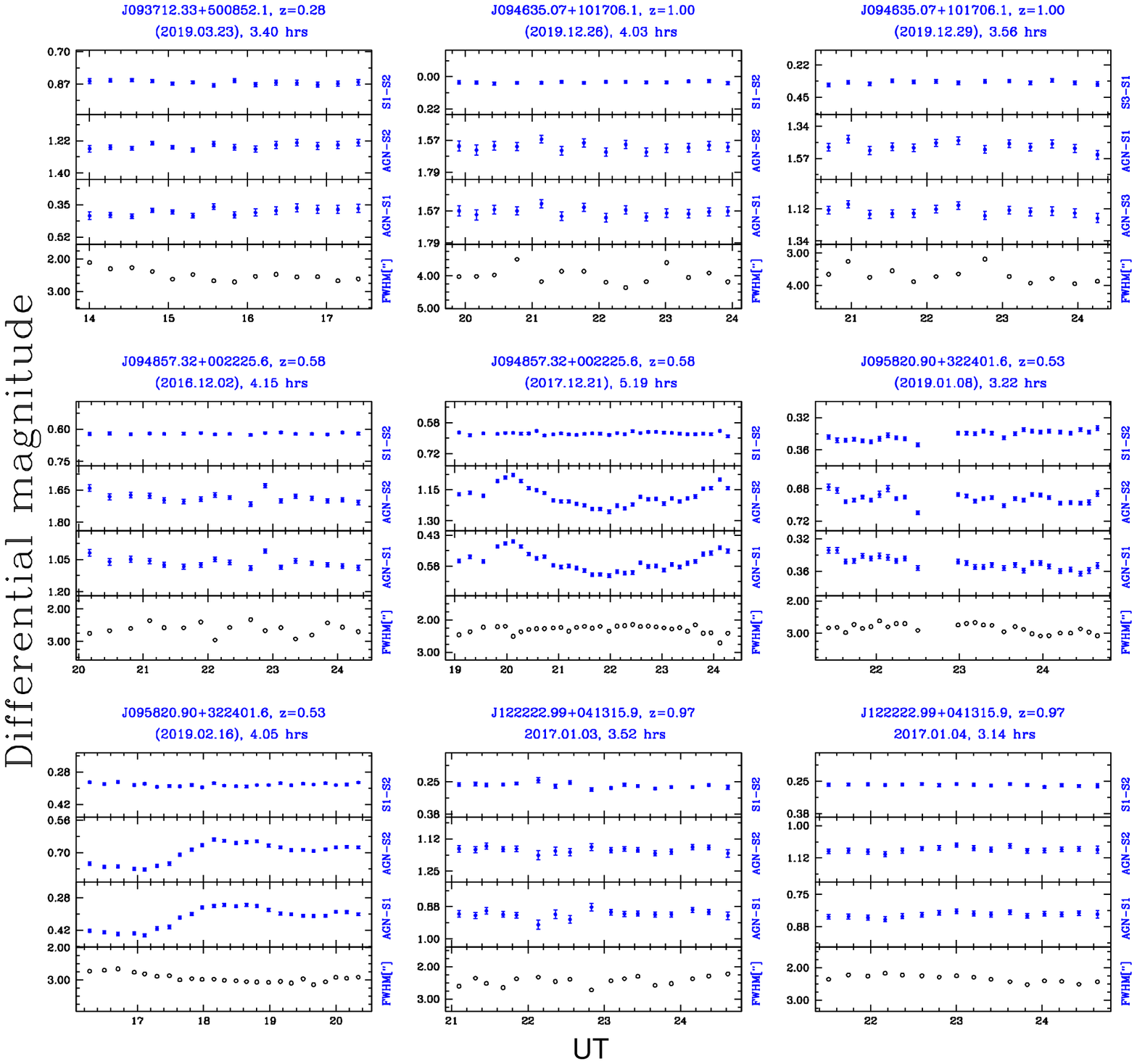}
\caption[]{Same as Fig.~\ref{fig:lurve 1}, but for the next 5 $\gamma-$ray NLSy1s from our 
sample of 15 $\gamma-$ray NLSy1s galaxies.}
\label{fig:lurve 2}
\end{figure*}
\begin{figure*}
\centering
\epsfig{figure=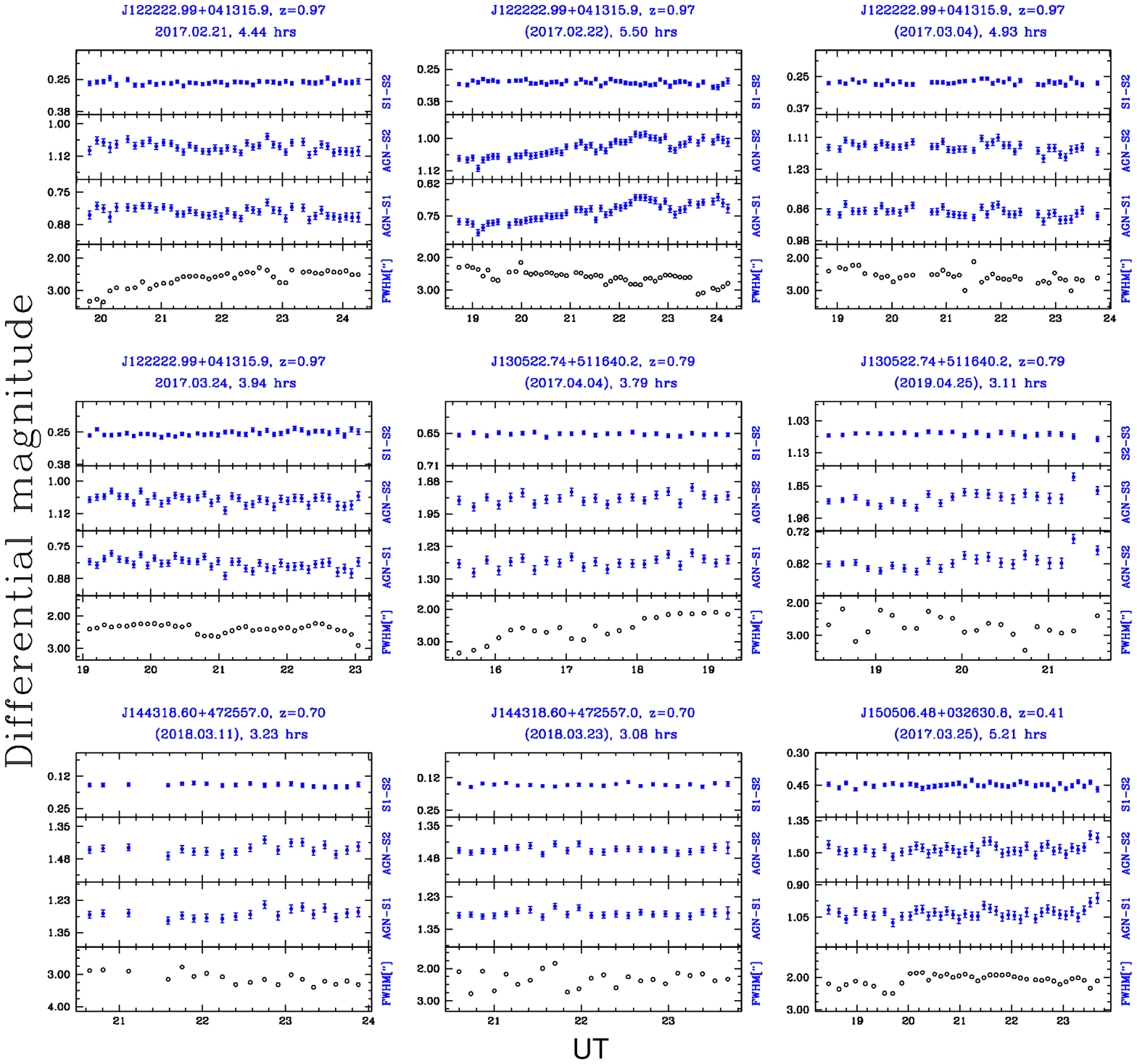}
\caption[]{Same as Fig.~\ref{fig:lurve 1}, but for the next 3 $\gamma-$ray NLSy1s from 
our sample of 15 $\gamma-$ray NLSy1s galaxies.}
\label{fig:lurve 3}
\end{figure*}
\begin{figure*}
\epsfig{figure=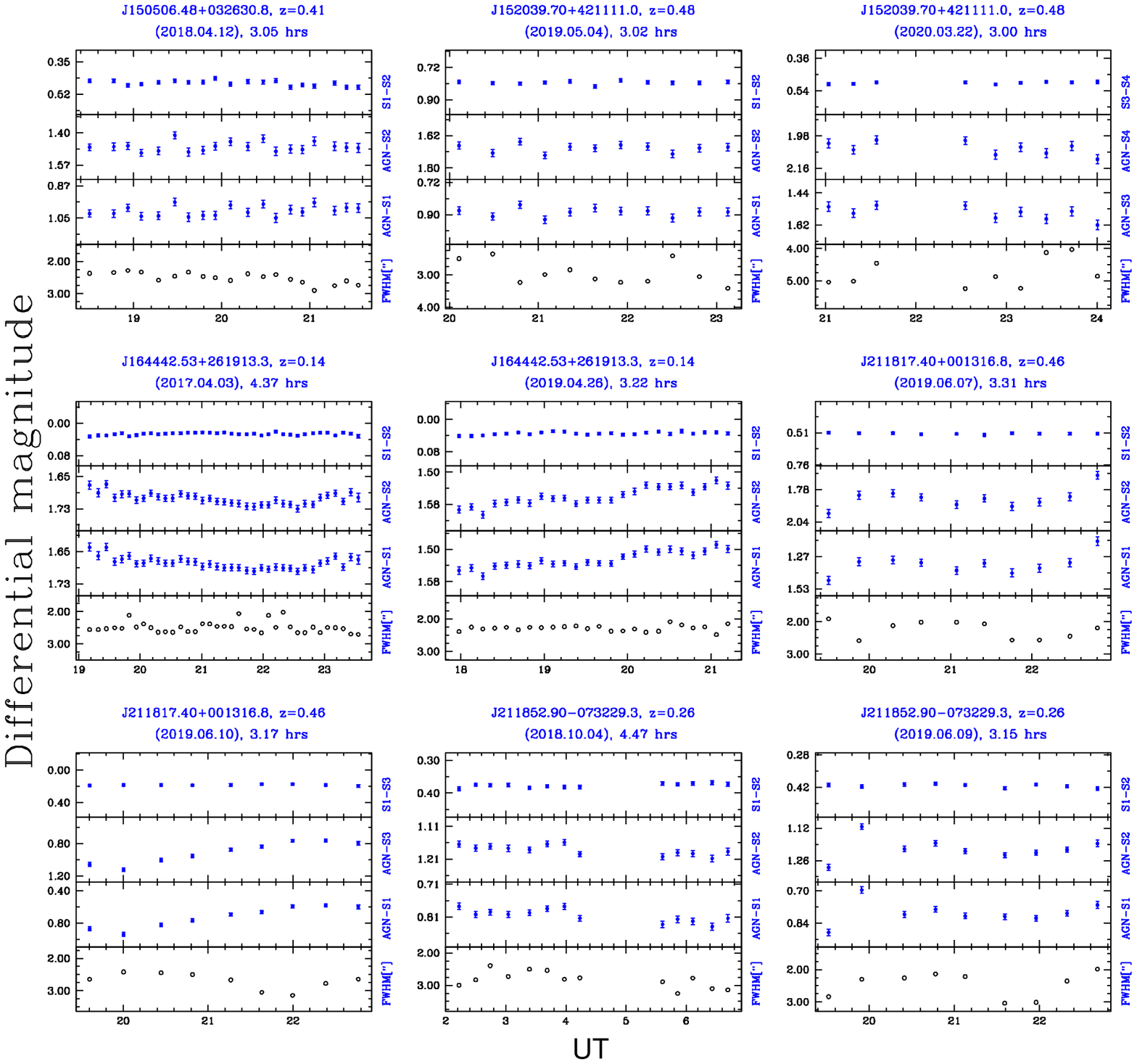}
\caption[]{Same as Fig.~\ref{fig:lurve 1}, but for the last 4 $\gamma-$ray NLSy1s 
from our sample of 15 $\gamma-$ray NLSy1s galaxies.}
\label{fig:lurve 4}
\end{figure*}

From Fig.~\ref{fig:lurve 1}, one of the two events is associated, once again, with J032441.20$+$341045.0 ($z$ = 0.06), when roughly in the middle of its 3.4 hrs long monitoring session on 2017-01-04, i.e., around 15:00 UT, its flux underwent a sharp drop by $\sim$ 3\% within $\sim$ 6 minutes and then, after remaining stagnant for $\sim$ 20 minutes, the flux jumped back by $\sim$ 3\% within 6 minutes. Now, following the arguments made by us previously in ~\citet{Ojha2019MNRAS.483.3036O} and recalled earlier in this section, the actual amplitude of both these flux changes are likely to be several times greater once a correction is made for the dilution of the jet's optical emission, due to the (much steadier) thermal emission  contributed by this strongly accreting AGN itself and also by its host galaxy. It is also interesting to note that out of the total 10 monitoring sessions of $>$ 3 hr duration, reported for this AGN here (5 sessions) and in previous studies~\citep{Paliya2013MNRAS.428.2450P, Paliya2014ApJ...789..143P, Ojha2019MNRAS.483.3036O}, strong features of duration $\ll$ 1 hr have been observed in the DLCs of as many as 3 sessions. Thus, large and ultra-fast brightness changes appear to be a remarkable behavioural signature of this $\gamma$-ray NLSy1 galaxy! Moving now to the second large event of comparable sharpness, found  in the present dataset, it can be seen in the DLCs of the source J094857.32+002225.8 ($z$ = 0.584) on 2017-12-21. During that 5.2 hrs long monitoring session  (Fig.~\ref{fig:lurve 2}), at around 19.55 UT, the flux showed a sharp jump (between two consecutive points) of $\sim$ 7\% within $\sim$ 15 minutes. Although, for this source, no estimates are currently available for the dilution by thermal optical emission, such a contamination is quite likely (especially from the AGN accretion disc), going by the above-mentioned example of the $\gamma$-ray NLSy1 J032441.20$+$341045.0. Note that a very strong INOV of J094857.32+002225.8  has been reported previously in the independent campaigns conducted by~\citet{Liu2010ApJ...715L.113L} and~\citet{Maune2013ApJ...762..124M}. The most rapid flux evolution, found by the latter team, occurred on 2011-04-01 and amounted to a rate of 0.2 - 0.3 mag hr$^{-1}$. Thus, both these  $\gamma$-ray NLSy1s are proto-types of blazar-like jet activity, probably even surpassing them in the rapidity of brightness change. It is also interesting to note that  these two AGN, although grossly different in redshift, display highly superluminal nuclear radio jets, with a speed of up to 7c in case of J032441.20$+$341045.0~\citep{Fuhrmann2016RAA....16..176F} and  9c for J094857.32$+$002225.6~\citep{Lister2019ApJ...874...43L}. Thus, in summary, the fact that in the present large and unbiased sample of $\gamma$-ray NLSy1s, at least 2 out of total 29 sessions have an event with optical flux doubling time of $\lesssim$ 1 hr. In contrast,  such rapid optical variations have hardly ever been observed  in the extensive monitoring programs targetting blazars (see above). Could the contrast be a reflection of the qualitative physical differences between the central engines of these two types of jetted AGNs (Sect.~\ref{sec1.0})? This appears to be a potentially important emerging clue, to be probed by further observations.

\section{Conclusions}
\label{sect 6.0}

Although the first INOV observations of a $\gamma$-ray NLSy1 galaxy were reported a decade ago~\citep{Liu2010ApJ...715L.113L}, subsequent follow up has been limited to just 6 members of this intriguing class of AGN. The present study, based on a much larger sample (see below) is the first attempt to systematically characterise the INOV properties of $\gamma$-ray NLSy1 galaxies. These jetted AGNs are particularly interesting since their central engines are thought to operate in a physical regime of accretion rate different from what is believed to occur in (more powerful) blazars. Our study is based on an unbiased sample of 15 $\gamma$-ray NLSy1s, assembled by applying well-defined selection criteria to a total of 20 such objects known at present. Based on their monitoring in 36 sessions of minimum 3 hrs duration, we estimate an INOV duty cycle of 25 - 30 per cent for a typical INOV amplitude detection threshold of around 3 - 5\%.  Among the powerful AGNs, only blazars are known to achieve INOV DC exceeding $\sim$ 20\%. Thus, as a class, $\gamma$-ray NLSy1s resemble blazars even in INOV properties. As discussed in Sect.~\ref{sec 5.0}, it is quite likely that we may have underestimated the DC for $\gamma$-ray NLSy1s because their nonthermal optical emission which is primarily responsible for INOV, could be significantly diluted by the much less variable optical emission contributed by the host galaxy (in case of lower redshift sources) and the accretion disc operating at a high Eddington accretion rate~\citep{Boroson1992ApJS...80..109B, Pounds1995MNRAS.277L...5P, Sulentic2000ApJ...536L...5S, Boroson2002ApJ...565...78B, Collin2004A&A...426..797C, Grupe2004ApJ...606L..41G, Paliya2019JApA...40...39P, Ojha2020ApJ...896...95O}.\par  
Further, we have detected in the light curves of two prominent members of our sample of 15 $\gamma$-ray NLSy1s, ultra-rapid brightness changes amounting to a clear level change within $\sim$ 0.1 hr. Quite plausibly, with a similarly dense sampling of the light curves for our entire sample, more such events would be found. But already, this result seems to stand in contrast to the observed extreme rarity of such distinct ultra-rapid events of optical variability in blazar light curves, even though they vastly outnumber the light curves currently available for $\gamma$-ray NLSy1s. If this difference is firmly established by further observations of $\gamma$-ray NLSy1s, that would provide a useful input to the models of jet formation in these enigmatic AGNs.\par

\section*{Acknowledgements}
We thank an anonymous referee for his/her very important comments that helped us to improve this manuscript considerably. We also thank the concerned ARIES staff for assistance during the observations. G-K thanks the Indian National Science Academy for a Senior Scientist fellowship.

\section*{Data availability}

The data used in this article are from 1.3-m Devasthal Fast Optical Telescope (DFOT) of Aryabhatta Research Institute of Observational Sciences (ARIES), Nainital, India (https://www.aries.res.in/facilities/astronomical-telescopes/13m-telescope), which will be shared on reasonable request to the corresponding author.

\bibliography{references}
 \label{lastpage}
 \end{document}